\documentclass[prl,superscriptaddress,twocolumn,footinbib]{revtex4-1}
\pdfoutput=1
\usepackage[colorlinks=true,urlcolor=blue]{hyperref}
\usepackage[normalem]{ulem}
\usepackage{mathtools}
\usepackage{amsfonts}
\usepackage{graphicx}
\usepackage{stmaryrd}
\usepackage{amsmath}
\usepackage{amssymb}
\usepackage{lineno}
\usepackage{xcolor}
\usepackage{color}
\usepackage{bbold}
\usepackage{bm}

\newcommand{\stkout}[1]{\ifmmode\text{\sout{\ensuremath{#1}}}\else\sout{#1}\fi}

\DeclareMathOperator{\tr}{tr}
\DeclareMathOperator{\Arg}{Arg}
\DeclareMathOperator{\sign}{sign}

\newcommand{\traceless}[1]{\left\llbracket #1 \right\rrbracket}
\newcommand{\OPeq}{|\Psi_{0}|}
\newcommand{\OP}{|\Psi|}
\newcommand{\deps}{\dot{\epsilon}}
\newcommand{\Hp}{\mathfrak{H}_{p}}

\newcommand{\er}{{\rm Er}}
\newcommand{\D}{\mathcal{D}}
\newcommand{\De}{\D_{\rm eff}}
\newcommand{\Ge}{\gamma_{\rm eff}}

\newcommand{\U}{\mathfrak{U}}
\newcommand{\ls}{\ell_{\rm s}}

\definecolor{red}{rgb}{0.75,0,0}
\definecolor{blue}{rgb}{0,0,0.75}
\definecolor{green}{rgb}{0,0.5,0}

\newcommand{\cred }{\color{red}}

\begin{document}
	

\title{Long-ranged order and flow alignment in sheared $p-$atic liquid crystals}

\author{Luca Giomi}
\email{giomi@lorentz.leidenuniv.nl}
\affiliation{Instituut-Lorentz, Universiteit Leiden, P.O. Box 9506, 2300 RA Leiden, The Netherlands}

\author{John Toner}
\affiliation{Department of Physics and Institute of Theoretical Science,  University of Oregon, Eugene, Oregon 97403, USA}

\author{Niladri Sarkar}
\affiliation{Instituut-Lorentz, Universiteit Leiden, P.O. Box 9506, 2300 RA Leiden, The Netherlands}

\begin{abstract}
We formulate a hydrodynamic theory of $p-$atic liquid crystals, namely two-dimensional anisotropic fluids endowed with generic $p-$fold rotational symmetry. Our approach, based on an order parameter tensor that directly embodies the discrete rotational symmetry of $p-$atic phases, allows us to unveil several unknown aspects of flowing $p-$atics, that previous theories, characterized by ${\rm O(2)}$ rotational symmetry, could not account for. This includes the onset of long-ranged orientational order in the presence of a simple shear flow of arbitrary shear rate, as opposed to the standard quasi-long-ranged order of two-dimensional liquid crystals, and the possibility of flow alignment at large shear rates.
\end{abstract}

\maketitle

One of the most surprising results in condensed matter physics is the prediction by Halperin and Nelson~\cite{Halperin:1978,Nelson:1979} that  two-dimensional solids can melt via a two-step process, known as the Kosterlitz-Thouless-Halperin-Nelson-Young (KTHNY) scenario \cite{Kosterlitz:1972,Kosterlitz:1973,Young:1979}. Upon increasing temperature, the unbinding of neutral dislocation pairs transforms a two-dimensional crystal, characterized by quasi-long-ranged translational order and long-ranged $p-$fold orientational order (i.e. invariance with respect to rotations by $2\pi/p$, with $p=1,\,2,\,3\ldots$), into a $p-$atic liquid crystal, in which the orientational order is preserved, albeit reduced to quasi-long-ranged, while translational order is lost. A further increase in temperature, drives the transition of the liquid crystal into an isotropic liquid, where both translational and orientational order are short-ranged order. As crystals comprising isotropically interacting building blocks are generally sixfold coordinated, the KTHNY scenario implies the existence of a hexatic phase (i.e. $p=6$) intermediate between two-dimensional crystals and isotropic liquids.

Since its theoretical prediction, $p-$atic phases and KTHNY melting scenario have been subject to extensive theoretical and experimental investigation~\cite{Bladon:1995,Bernard:2011, Zahn:1999,Gasser:2010,Thorneywork:2017,Anderson:2017}). By contrast, the hydrodynamic behavior of $p-$atics has received little attention and, with the exception of a small number of pioneering works, e.g. Refs. \cite{Zippelius:1980a, Zippelius:1980b, Sonin:1998, Krieger:2014}, is still largely unexplored. Furthermore, previous hydrodynamic theories of $p-$atics are characterized by ${\rm O}(2)$ rotational symmetry, which is higher than the actual $p-$fold symmetry of $p-$atic phases. 

Yet, recent findings in tissue mechanics have reignited interest in $p-$atic hydrodynamics, by providing these phases of matter with unexpected biological relevance. Using a popular cell-resolved model of confluent epithelial tissues \cite{Nagai:2001,Farhadifar:2007}, Li and Picaciamarra have recently demonstrated that, as for two-dimensional crystals, the solid and the isotropic liquid states of these model-epithelia are separated by an intermediate hexatic phase, in which cells are orientationally ordered and yet able to flow \cite{Li:2018}. This remarkable discovery sheds new light on the complex physics of tissues and, simultaneously, provides a strong motivation for investigating hexatic hydrodynamics  and, more generally, the hydrodynamics of $p-$atic liquid crystals, more deeply.

In this Letter, we go beyond the classic ${\rm O}(2)$ picture of $p-$atic hydrodynamics. Using a phenomenological approach rooted in the $p-$atic tensor order parameter, whose algebraic structure directly embodies the discrete rotational symmetry of $p-$atics, we identify additional couplings between $p-$atic order and flow. These novel couplings leave a distinct signature on the high shear rate dynamics, which may cause  the $p-$atic director to align at specific system-dependent angles with respect to the underlying velocity field. Moreover, we demonstrate that a shear flow of arbitrary finite shear rate has the remarkable effect of turning quasi-long-ranged orientational order, i.e. the hallmark of two-dimensional liquid crystals at equilibrium, into long-ranged order. A longer and more detailed account of our results is given in Ref. \cite{Giomi:2021}.

\begin{figure}[t]
\centering
\includegraphics[width=\columnwidth]{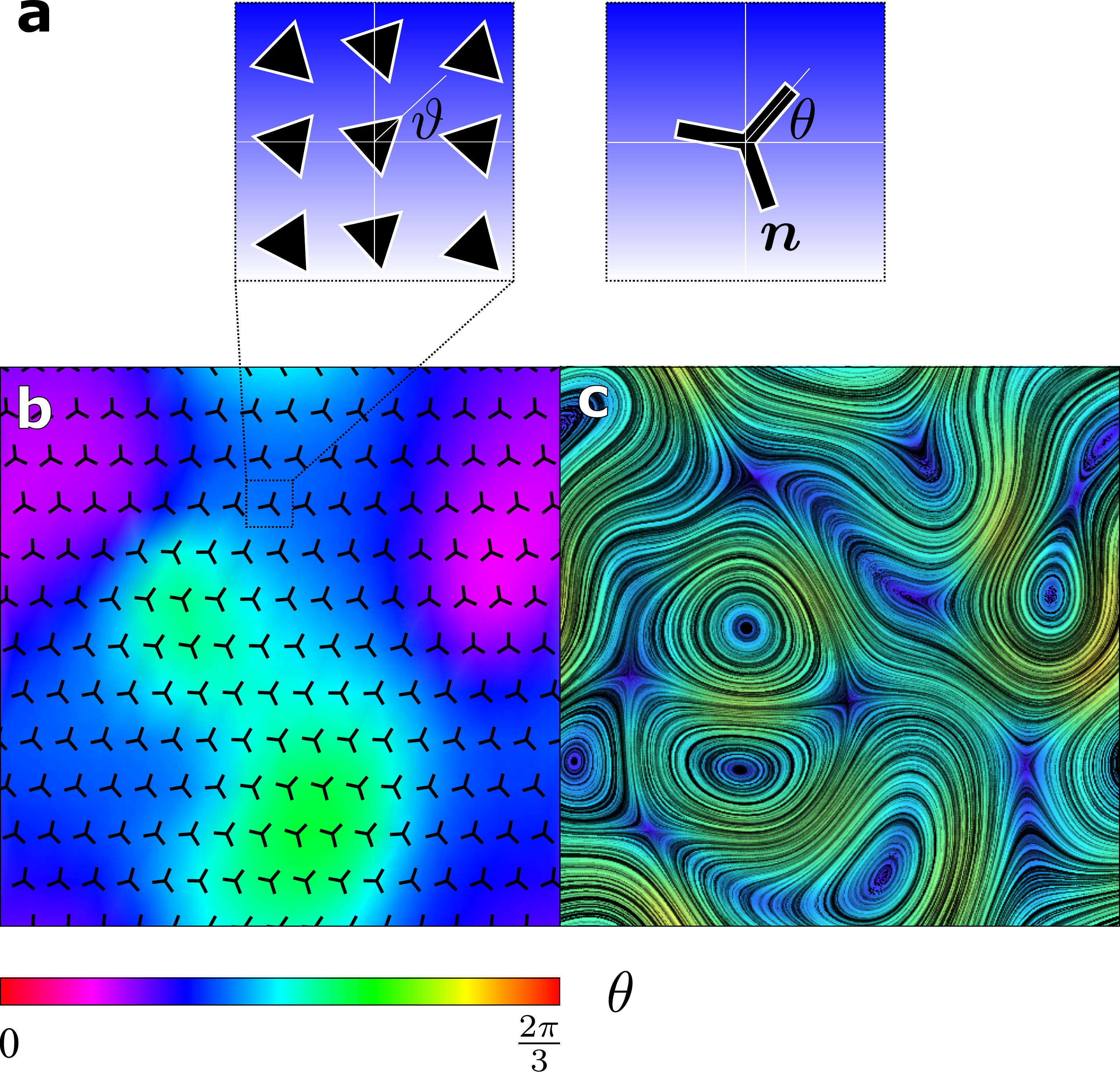}
\caption{\label{fig:snapshots}(a) Illustration of triatic building blocks (left) together with the corresponding coarse-grained $p-$atic director (right). (b,c) Typical configuration of the triatic director (b) and velocity field (c) coarsening from an initially disordered state. The data in displayed in panels (b) and (c) have been obtain by a numerical integration Eqs. \eqref{eq:hydrodynamics} in the incompressible limit (i.e. $\rho={\rm const}$).}	
\end{figure}

We consider a generic $p-$atic liquid crystal, whose microscopic constituents can be assigned a direction $\bm{\nu}=\cos\vartheta\,\bm{e}_{x}+\sin\vartheta\,\bm{e}_{y}$ (Fig. \ref{fig:snapshots}a). The latter may represent a specific molecular direction, corresponding e.g. to a particular functional group, or be conventionally assigned. Local $p-$atic order is then embodied in the complex function $\psi_{p}=e^{ip\vartheta}$, whose correlation function decays algebraically in equilibrium: i.e. $\left\langle \psi_{p}^{*}(\bm{r})\psi_{p}(\bm{0}) \right\rangle \sim |\bm{r}|^{-\eta_{p}}$, where $0<\eta_{p}\le 1/4$ is a non-universal (i.e., temperature-dependent) exponent \cite{Halperin:1978,Nelson:1979}. Equivalently, the $p-$atic order parameter
\begin{equation}\label{eq:p-atic_order_parameter}
\Psi_{p} = \left\langle	\psi_{p} \right\rangle = |\Psi| e^{ip\theta}\,,
\end{equation}
with $\theta$ the average molecular orientation (Fig. \ref{fig:snapshots}a), depends on the length scale $\ell$ at which it is probed, i.e. $|\Psi|\sim \ell^{-\eta_{p}/2}$, and vanishes in the infinite system size limit: i.e. $\lim_{\ell\to\infty}\OP=0$. 

The central object in our approach is the rank$-p$ tensor order parameter, $\bm{Q}_{p}=Q_{i_{1}i_{2}\cdots\,i_{p}}\bm{e}_{i_{1}}\otimes\bm{e}_{i_{2}}\otimes\cdots
\otimes\bm{e}_{i_{p}}$ with $i_{n}=\{x,y\}$ and $n=1,\,2\ldots\,p$, constructed upon averaging the $p-$th tensorial power of the local orientation $\bm{\nu}$. That is
\begin{equation}\label{eq:p-atic_tensor}
\bm{Q}_{p} = \sqrt{2^{p-1}} \traceless{\langle \bm{\nu}^{\otimes p} \rangle} = \sqrt{2^{p-1}}|\Psi|\traceless{\bm{n}^{\otimes p}}\,,	
\end{equation}
where $\bm{n}=\cos\theta\,\bm{e}_{x}+\sin\theta\,\bm{e}_{y}$ and the operator $\traceless{\cdots}$ has the effect of rendering an arbitrary tensor symmetric with respect to the exchange of any two indices and traceless, i.e. $Q_{jji_{3}\dots\,i_{p}}=0$ \cite{Hess:2015}. For $p=2$, Eq. \eqref{eq:p-atic_tensor} readily gives the standard nematic order parameter tensor: i.e. $\bm{Q}_{2}=|\Psi|(\bm{n}\otimes\bm{n}-\mathbb{1}/2)$, with $\mathbb{1}$ the identity tensor. 

Together with the mass density $\rho$ and the momentum density $\rho\bm{v}$, with $\bm{v}$ the local velocity field, the order parameter tensor $\bm{Q}_{p}$ is a ``hydrodynamic variable" of $p-$atics, that is,  a material field whose relaxation rate vanishes at large length scales~\cite{Forster:1975}. The hydrodynamic equation governing the spatiotemporal evolution of such a broken symmetry variable can be obtained by expressing its time derivative as a sum of all possible symmetric and traceless rank$-p$ tensor combinations of the velocity gradient tensors $\nabla\bm{v}$ with $\bm{Q}_{p}$ and its gradients. This procedure, detailed in Ref. \cite{Giomi:2021}, yields the following set of partial differential equations
\begin{subequations}\label{eq:hydrodynamics}
\begin{gather}
\frac{D\rho}{Dt}+\rho\nabla\cdot\bm{v} = 0\,,\\
\rho\frac{D\bm{v}}{Dt} = \nabla\cdot\bm{\sigma}+\bm{f}\,,\\
\frac{D\bm{Q}_{p}}{Dt} = \Gamma\bm{H}_{p} + p \big \llbracket \bm{Q}_{p}\cdot\bm{\omega} \big \rrbracket + \bar{\lambda}_{p}\tr(\bm{u})\bm{Q}_{p} \notag \\
+ \lambda_{p} \big\llbracket \nabla^{\otimes p-2}\bm{u} \big\rrbracket + \nu_{p} \big \llbracket \nabla^{\otimes p\,{\rm mod}\,2} \bm{u}^{\otimes\lfloor p/2 \rfloor} \big \rrbracket\,,
\end{gather}	
\end{subequations}
where $D/Dt=\partial_{t}+\bm{v}\cdot\nabla$ is the material derivative, the rank$-2$ tensors $\bm{\omega}=[\nabla\bm{v}-(\nabla\bm{v})^{\rm T}]/2$ and $\bm{u}=[\nabla\bm{v}+(\nabla\bm{v})^{\rm T}]/2$, with ${\rm T}$ indicating transposition, are, respectively, the vorticity and strain rate tensors, and the dot product implies a contraction of one index on$\bm{Q}$ with one on $\bm{\omega}$. That is: $(\bm{Q}_{p}\cdot\bm{\omega})_{i_{1}i_{2}\dots\,i_{p}}=Q_{i_{i}i_{2}\dots\,j}\omega_{ji_{p}}$.

Eqs. (\ref{eq:hydrodynamics}a) and  (\ref{eq:hydrodynamics}b) are the standard mass and momentum conservation equations, with $\bm{\sigma}$ the stress tensor and $\bm{f}$ an arbitrary body force. In Eq. (\ref{eq:hydrodynamics}c), $\Gamma^{-1}$ is a rotational viscosity and $\bm{H}_{p}=-\delta F/\delta \bm{Q}_{p}$ is the molecular tensor describing the relaxation of the $p-$atic phase toward the minimum of the free energy $F$. This, in turn, can be constructed in the standard Landau-de Gennes form \cite{DeGennes:1993}:
\begin{equation}
F = \int {\rm d}A\,\left(\frac{1}{2}\,L|\nabla\bm{Q}_{p}|^{2}+\frac{1}{2}\,a_{2}|\bm{Q}_{p}|^{2}+\frac{1}{4}\,a_{4}|\bm{Q}_{p}|^{4}\right)\,, 
\end{equation}
where $|\cdots|^{2}$ indicates for the Euclidean norm, obtained from the full contraction of a tensor with itself. In particular: $|\bm{Q}_{p}|^{2}=|\Psi|^{2}/2$. The constant $L$ is the order parameter stiffness, while $a_{2}$ and $a_{4}$ are phenomenological coefficients favoring a non-vanishing $|\Psi|$ value in the ordered phase, where $a_{2}<0$. Specifically, $|\Psi|=|\Psi_{0}|=\sqrt{-2a_{2}/a_{4}}$ at the minimum of the free energy. On the second line of Eq. (\ref{eq:hydrodynamics}c), $(\nabla^{\otimes n})_{i_{1}i_{2}\cdots\,i_{n}}=\partial_{i_{1}}\partial_{i_{2}}\cdots\,\partial_{i_{n}}$, while $\lfloor \ldots \rfloor$ denotes the floor function and $p\,\mod\,2=p-2\lfloor p/2 \rfloor$ is zero for even $p$ values and one for odd $p$ values. Finally, $\bar{\lambda}_{p}$, $\lambda_{p}$ and $\nu_{p}$ are material parameters expressing the strength of the coupling between $p-$atic order and flow.

The stress tensor $\bm{\sigma}$ can be routinely decomposed in a viscous (i.e. energy dissipating) and a reactive (i.e. energy conserving) contribution: i.e. $\bm{\sigma}=\bm{\sigma}^{({\rm v})}+\bm{\sigma}^{({\rm r})}$. Following Onsager's ’s reciprocal relations the former can be expressed as $\bm{\sigma}^{({\rm v})}=\bm{\eta}:\nabla\bm{v}$, where the colon product denotes a contraction of two indices and $\bm{\eta}$ is the rank$-4$ viscosity tensor. As  in conventional liquid crystals, this tensor could have, {\em a priori}, both an isotropic and an anisotropic component: i.e. $\bm{\eta}=\bm{\eta}^{({\rm i})}+\bm{\eta}^{({\rm a})}$, with 
\begin{equation}
\eta_{ijkl}^{({\rm i})}=\zeta\delta_{ij}\delta_{kl}+\eta(\delta_{ik}\delta_{jl}+\delta_{il}\delta_{jk}-\delta_{ij}\delta_{kl})\,,
\end{equation}
and $\zeta$ and $\eta$ the bulk and shear viscosity respectively. However, for all $p>2$, {\em except} $p=4$, as shown in Ref.~\cite{Giomi:2021}, $\bm{\eta}^{({\rm a})}$ vanishes. Hence, kinetic energy is isotropically dissipated throughout the flow. For $p=4$, on the other hand, $\bm{\eta}^{({\rm a})}=\varrho_{4}\traceless{\bm{n}^{\otimes 4}}$, where $-2\eta \le \varrho_{4} \le 2\eta$ is an anisotropic viscosity, and the inequalities follow from the second law of thermodynamics \cite{Giomi:2021}. Similarly, calculating the entropy production rate allows one to express the reactive stress tensor in the form $\bm{\sigma}^{({\rm r})}=-P\mathbb{1}+\bm{\sigma}^{({\rm e})}+\bm{\sigma}^{({\rm d})}$, where 
\begin{equation}\label{eq:}
\sigma_{ij}^{({\rm e})}=-L\partial_{i}Q_{k_{1}k_{2}\cdots\,k_{p}}\partial_{j}Q_{k_{1}k_{2}\cdots\,k_{p}}\,,
\end{equation}
is the {\em elastic} stress, arising in response of a static deformation of a fluid patch, and
\begin{align}\label{eq:dynamic_stress}
\sigma_{ij}^{({\rm d})} 
&=- \bar{\lambda}_{p} Q_{k_{1}k_{2}\cdots\,k_{p}}H_{k_{1}k_{2}\cdots\,k_{p}}\,\delta_{ij} \phantom{\frac{p}{2}} \notag \\
&+ (-1)^{p-1}\lambda_{p}\partial^{p-2}_{k_{1}k_{2}\cdots\,k_{p-2}}H_{k_{1}k_{2}\cdots\,ij} \notag \\
&+ \frac{p}{2} \left( Q_{k_{1}k_{2} \cdots\,i} H_{k_{1}k_{2}\cdots\,j}-H_{k_{1}k_{2}\cdots\,i}Q_{k_{1}k_{2}\cdots\,j} \right)\,,
\end{align}	
is the {\em dynamic} stress originating from the reversible coupling between $p-$atic order and flow. 

Lastly, in the absence of topological defects or other singular features, the phase $\theta$ of the complex order parameter is the only hydrodynamic variable arising from the broken rotational symmetry, whereas $\OP$ relaxes to its equilibrium value in a finite time. In this case, Eq. (\ref{eq:hydrodynamics}c) can be simplified the form
\begin{equation}\label{eq:theta_p}
\frac{D\theta}{Dt} 
= \frac{K}{\gamma}\nabla^{2}\theta+\omega_{xy}
-|\Hp|\sin\left(p\theta-\Arg\Hp\right)\,,
\end{equation}
where the constants $K$ and $\gamma$ are related to $L$ and $\Gamma$ by 
\begin{equation}
K = \frac{p^{2}\OP^{2}}{2}\,L\,, \qquad \gamma = \frac{p^{2}\OP^{2}}{2}\,\Gamma^{-1}\,,	
\end{equation}
whereas the complex function
\begin{equation}
\Hp = \frac{2}{p\OPeq}\left(\lambda_{p}\partial^{p-2}\U+\nu_{p}\partial^{p\,{\rm mod}\,2}\U^{\lfloor p/2 \rfloor}\right)\,,
\end{equation}
with $\U=(u_{xx}-u_{yy})/2+iu_{xy}$, embodies the interplay between $p-$atic order and flow and will be hereafter referred to as {\em flow alignment field}.

Together with a system-specific equation of state, relating pressure and density (e.g. $P=c_{\rm s}^{2}\rho$, with $c_{\rm s}$ the speed of sound), Eqs. \eqref{eq:hydrodynamics} and the stress tensors given above govern the dynamics of a generic $p-$atic liquid crystal with $p \ge 2$, subject to arbitrary external forcing. The case $p=1$ requires a separate treatment and is discussed in Ref. \cite{Giomi:2021}. 

Before illustrating specific examples of viscous flow in $p-$atics, some remarks are in order. Eq. (\ref{eq:hydrodynamics}c) implies that local $p-$atic order, embodied in the tensorial field $\bm{Q}_{p}$, evolves toward the free energy minimum, where $\bm{H}_{p}$ vanishes, while simultaneously interacting with flow. Contrary to assertions in earlier literature, however, this interaction is not limited to the precession of the director $\bm{n}$ in the vorticity field, but includes couplings with the local strain rate. Among these, the term proportional to $\bar{\lambda}_{p}$ affects exclusively the scalar order parameter $|\Psi|$, whereas the terms proportional to $\lambda_{p}$ and $\nu_{p}$ also affect the local orientation $\theta$ and can drive reorientations of the $p-$atic director. In contrast with the cases of polar (i.e. $p=1$) and nematic (i.e. $p=2$) liquid crystals, however, these couplings depend on either derivatives or non-linear powers of the strain rate tensor, and are therefore expected to become important only at large shear rate or in rapidly spatially varying flows.

\begin{figure}[t!]
\centering
\includegraphics[width=\columnwidth]{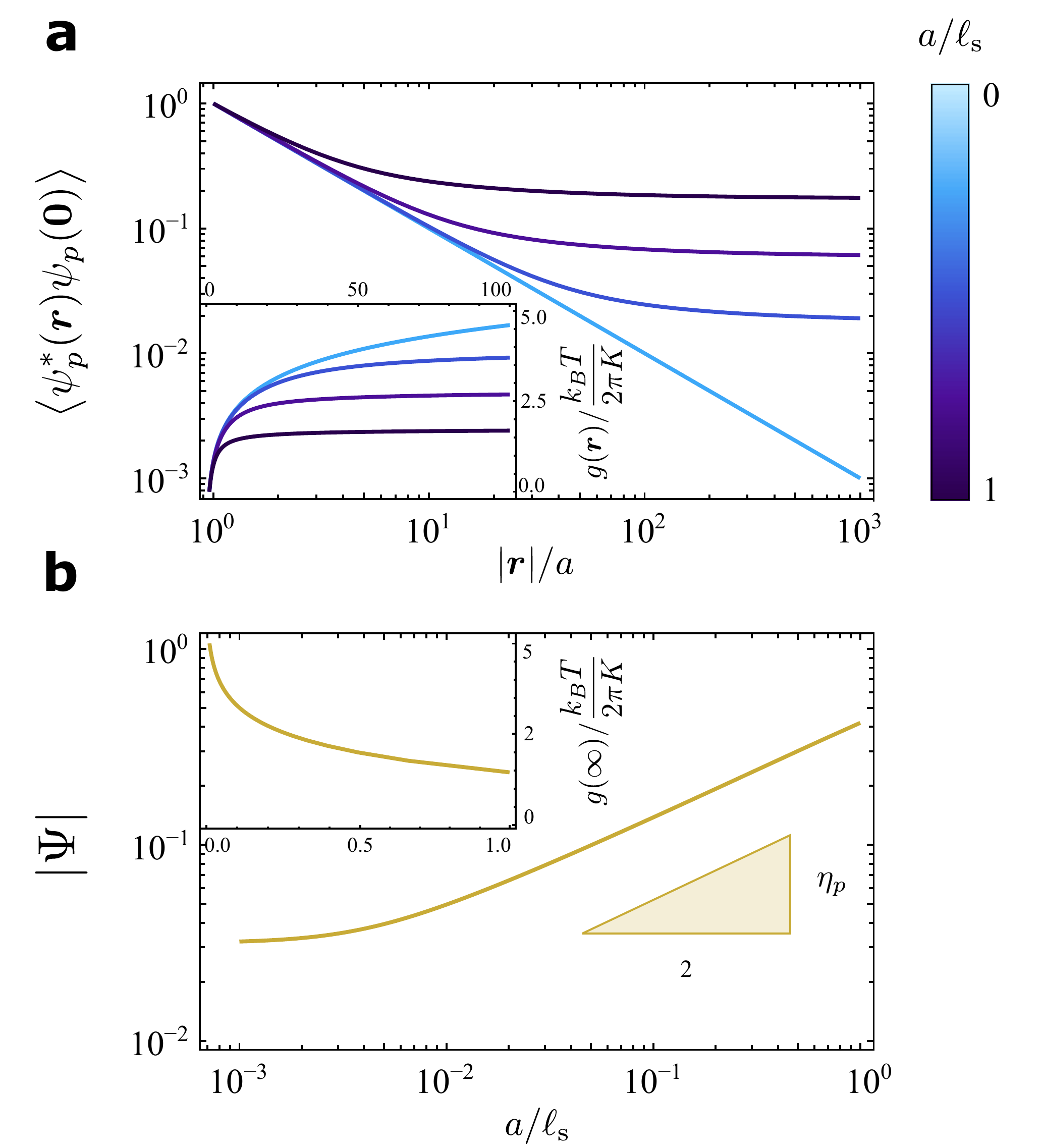}	
\caption{\label{fig:correlation_function} (a) Two-point $p-$atic correlation function versus distance for various shear rates expressed in terms of the dimensionless ratio $a/\ell_{\rm s}$, with $a$ a short distance cut-off and the shear length scale $\ell_{\rm s}$ for $\phi=0$. Inset: the connected correlation function $g=g(\bm{r})$, Eq. \eqref{eq:g}, versus distance. (b) $p-$atic order parameter $\OP$ versus shear rate, expressed in terms of $a/\ell_{\rm s}$. Inset: the asymptotic value $g(\infty) = \lim_{|\bm{r}|\rightarrow\infty}g(\bm{r})$.}
\end{figure}

With Eqs. \eqref{eq:hydrodynamics} in hand, we can now investigate the effect of a flow on $p-$atic order. To this end, we consider an incompressible system subject to an externally imposed shear flow, whose average velocity is given by $\langle \bm{v} \rangle = \deps y\bm{e}_{x}$. As demonstrated in Ref. \cite{Giomi:2021}, in the Stokesian limit (i.e. where inertial effects are negligible) and when $\rho K/\eta^{2}\ll 1$, a limit that applies extremely well to all known liquid crystals, Eqs. \eqref{eq:hydrodynamics} suitably supplemented by noise terms required by the fluctuation-dissipation theorem at finite temperature, can be cast in a single {\em linear} equation for the microscopic orientation $\vartheta$, i.e.:
\begin{equation}\label{eq:linear_eom}
\partial_{t}\vartheta+\deps y \partial_{x}\vartheta = \De\nabla^{2}\vartheta-\frac{\deps}{2}+\xi\,.
\end{equation}
The constant $\De=K[1/\gamma+1/(4\eta)]$ is an effective rotational diffusion coefficient, which further accounts for the internal {\em backflow} resulting from spatial variations of $p-$atic order, whereas $\xi$ a noise field, whose correlation function is given by 
\begin{equation}\label{eq:noise_correlation}
\left\langle\xi(\bm{r},t)\xi(\bm{r}',t')\right\rangle = \frac{2k_{\rm B}T}{\Ge}\,\delta(\bm{r}-\bm{r}')\delta(t-t')\,,
\end{equation}
with $\Ge=K/\De$. From the exact solution of Eq. \eqref{eq:linear_eom} \cite{Giomi:2021} one can then compute the correlation function of the complex $p-$atic order parameter in the form $\left\langle \psi_{p}^{*}(\bm{r})\psi_{p}(\bm{0}) \right\rangle = \exp[-p^{2}g(\bm{r})]$, with
\begin{equation}\label{eq:g}
g(\bm{r}) = \frac{k_{B}T}{2\pi K}\int_{0}^{\infty}{\rm d}\tau\,\frac{e^{-\mathcal{G}(\tau,\phi)\left(\frac{a}{\ls}\right)^{2}}-e^{-\mathcal{G}(\tau,\phi)\left(\frac{|\bm{r}|}{\ls}\right)^{2}}}{\tau\sqrt{4+\frac{1}{3}\tau^{3}}}\,,
\end{equation}
where
\begin{equation}\label{eq:phi}
\mathcal{G}(\tau,\phi) = \frac{1-\frac{1}{2}\tau\sin 2\phi+\frac{1}{3}\tau^{2}\sin^{2}\phi}{2\tau\left(4+\frac{1}{3}\tau^{2}\right)}\;,
\end{equation} 
$\phi=\arctan y/x$ and $\ls=\sqrt{\De/\deps}$, hereafter referred to as the {\em shear length scale}, is the distance at which elastic and hydrodynamic torques balance each other. Fig. \ref{fig:correlation_function}a (inset) shows a plot of $g(\bm{r})$ versus $|\bm{r}|/a$ for various $a/\ls$ values. For $a/\ls\rightarrow 0$, corresponding to $\deps \to 0$, this displays the characteristic logarithmic growth of $p-$atics at equilibrium: i.e. $g(\bm{r})=k_{B}T/(2\pi K)\log |\bm{r}|/a$, from which $\eta_{p}=p^{2}k_{B}T/(2\pi K)$.  By contrast, for $a/\ell_{\rm s}>0$, $g(\bm{r})$ does not grow without bound, but rather plateaus at large distances. This implies that $\langle \psi_{p}^{*}(\bm{r})\psi_{p}(\bm{0})\rangle$ converges to a finite value (Fig. \ref{fig:correlation_function}a), indicating that a shear flow of arbitrary finite shear rate renders the orientational order of $p-$atic phases long-ranged. The corresponding order parameter $\OP$, can be calculated from the large distance limit of $\langle \psi_{p}^{*}(\bm{r})\psi_{p}(\bm{0})\rangle$ and is given by
\begin{equation}
\OP \sim \left(\frac{a}{\ls}\right)^{\eta_{p}/2} \sim \deps^{\,\eta_{p}/4}\,.
\label{LRO}	
\end{equation}
In summary, in the presence of a simple shear flow, fluctuations are anisotropic, as indicated by the $\phi-$dependence in Eq. \eqref{eq:phi}, but are suppressed at length scales larger than $\ls$, where the elastic torques, which alone would not suffice to break rotational symmetry, are overcome by hydrodynamic torques, resulting in the emergence of global alignment. Although this analysis ignores the nonlinear terms in Eqs. (\ref{eq:hydrodynamics}c), in Ref. \cite{Giomi:2021} we show that these terms are irrelevant in the Renormalization Group sense: i.e. they do not affect the long distance and time behavior of the system. Hence, the results quoted above, in particular Eq.~\eqref{LRO}, are valid at long length scales in real systems.

\begin{figure}[t]
\centering
\includegraphics[width=\columnwidth]{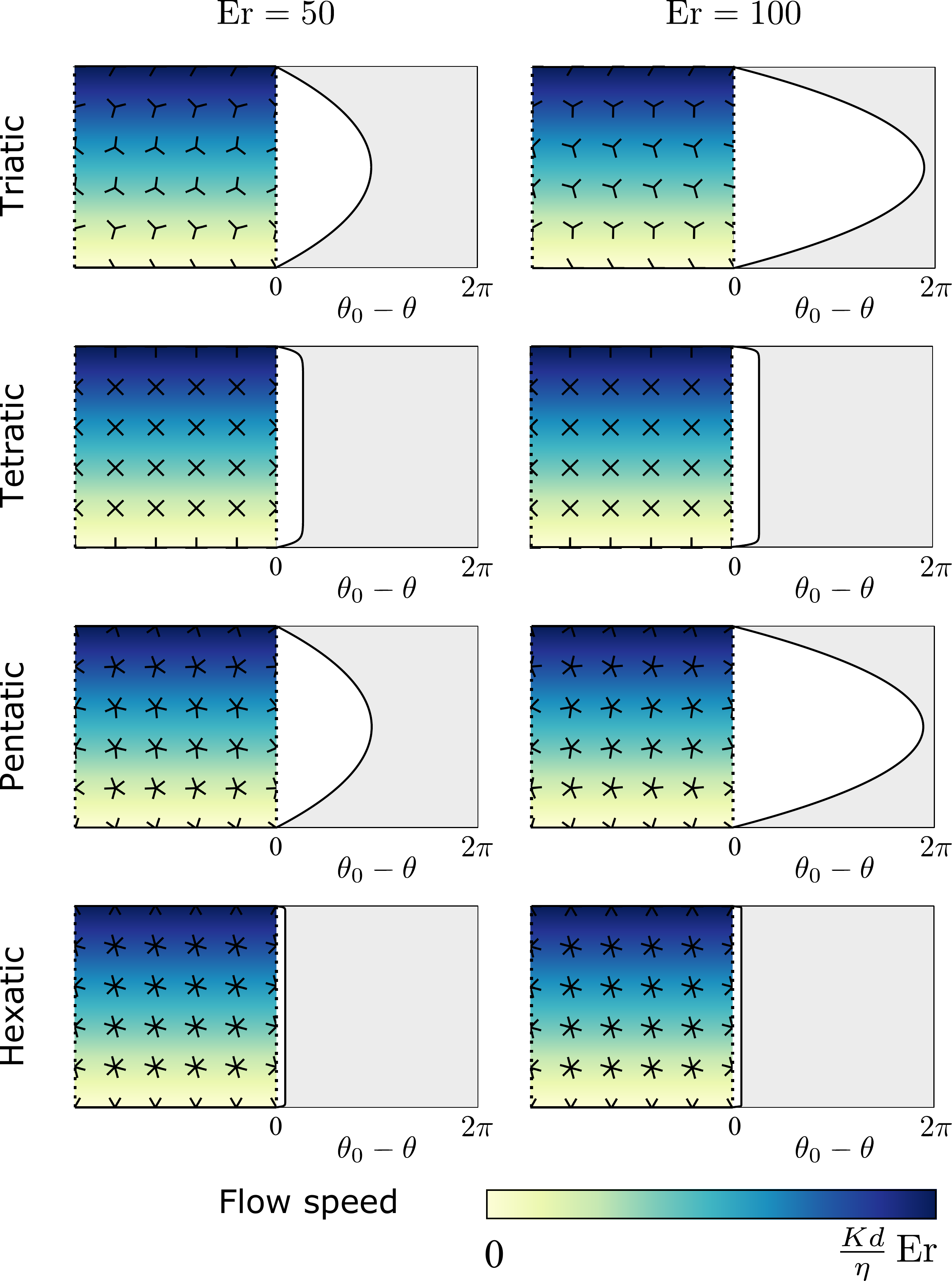}
\caption{\label{fig:channel}Examples of high Ericksen number $p-$atic flow in a channel. Numerical solution of Eq. \eqref{eq:theta_p} for triatics ($p=3$), tetratics ($p=4$), pentatics ($p=5$) and hexatics ($p=6$) with boundary conditions $\theta_{0}=\Delta\theta=0$, with $d$ the channel thickness. The left-hand side of all plots shows the configuration of the $p-$atic director, represented by $p-$headed stars, superimposed to a heat map of the flow speed. The solid lines denote the channel walls, whereas the dotted lines mark the position of the channel inlet/outlet. The left-hand side of the plots shows the configuration of the $p-$atic director in terms of the angle $\theta_{0}-\theta$. In all plots the parameter values are $\rho K/\eta^{2}=1$, $\lambda_{p}/d^{p-2}=1.5$ and $\nu_{p}/[(\rho/\eta)^{\lfloor p/2 \rfloor-1}d^{p-2}]=2.0$.}
\end{figure}

As a second example of the hydrodynamic behavior of $p-$atics, we consider a system confined within a two-dimensional channel of infinite length along the $x-$direction and finite width $d \gg a$ along the $y-$direction. At these large length scales, we may ignore thermal fluctuations. The upper wall is dragged at speed $v_{0}$, in such a way the so-called Ericksen number $\er=\eta v_{0}d/K$ (see e.g. Ref.~\cite{Kleman:2003}), expressing the preponderance of an externally induced flow with respect to the internal backflow, is large, so that the velocity field is unaffected by the orientational order, and therefore just what one would obtain in this geometry in a simple fluid; i.e. $\bm{v}=\deps y\bm{e}_{x}$, with $\deps=v_{0}/a$ a constant shear rate. Under these assumptions, a stationary configuration of the average orientation $\theta$ is found for $p=3,\,5,\,7\ldots$ by solving a simplified version of Eq. \eqref{eq:theta_p} given by
\begin{equation}\label{eq:shear_flow_odd}
\partial_{y}^{2}\theta = \frac{1}{2\ls^{2}}\,,
\end{equation}
with the shear length scale now given by $\ls=\sqrt{K/(\gamma\deps)}$ and whose solution with the boundary conditions $\theta(0)=\theta_{0}$ and $\theta(d)=\theta_{0}+\Delta\theta$ is
\begin{equation}\label{eq:theta_odd}
\theta(y) = \theta_{0}+\Delta\theta\,\frac{y}{d}+\frac{y(y-d)}{4\ls^{2}}\,.
\end{equation}
Thus, for odd $p \ne 1$, the director rotates in such a way to accommodate the vorticity of the imposed shear flow, as can be seen in Fig. \ref{fig:channel} in the case of triatics ($p=3$) and pentatics ($p=5$). As we demonstrate in Ref. \cite{Giomi:2021}, and as it is already known in nematics \cite{Thampi:2015}, this solution is unstable to a time-dependent or ``tumbling'' configuration, where the director periodically precesses across the channel, while temporarily disengaging from the boundary via a localized suppression of $p-$atic order. The instability occurs when the Ericksen number overcomes the threshold $\er_{\rm c}=(\eta/\gamma)(d/\xi_{\rm m})/(p/2)$, where $\xi_{\rm m}=\sqrt{L/|a_{2}|}$ is the $p-$atic coherence length. By contrast, for $p=4,\,5,\,6\ldots$, Eq. \eqref{eq:theta_p} reduces to 
\begin{equation}\label{eq:shear_flow_even}
\partial_{y}^{2}\theta = \frac{1}{2\ls^{2}}\left[1+\left(\frac{\deps}{\deps_{\rm c}}\right)^{p/2-1}\sin p\left(\theta-\frac{\pi}{4}\right)\right]\,,
\end{equation}
with $\deps_{\rm c}$ a constant shear rate given by
\begin{equation}\label{eq:critical_shear_rate}
\deps_{\rm c} = 2\left(\frac{p\OPeq}{2\nu_{p}}\right)^{\frac{1}{p/2-1}}\,. 
\end{equation}
For small $\deps$ value, the $p-$atic director rotates across the channel and its configuration is again approximatively describe by Eq. \eqref{eq:theta_odd}. Unlike in the case of odd $p$ values, however, increasing the shear rate does not trigger a flow tumbling instability. Conversely, the director aligns at an angle that progressively approaches the asymptotic value $\theta_{p}=(\pi/4+k\pi/p)\,{\rm mod}\,2\pi/p$, where the integer $k$ depends on the anchoring of the $p-$atic director and can be selected in such a way to minimize the energetic cost of the boundary layer in proximity of the channel walls. This yields: $\theta_{4}=\pm \pi/4$, $\theta_{6}=\pm \pi/12$, $\theta_{8}=\pm \pi/8$ etc. with the sign is given by $-\sign\dot{\epsilon}$. This is illustrated in Fig. \ref{fig:channel} for the cases of tetradics (i.e. $p=4$) and hexatics (i.e. $p=6$). The phenomenon is known ``flow alignment'' in the literature of liquid crystals, and was, so far, thought to occur exclusively in polar liquid crystals (i.e. $p=1$) and nematics (i.e. $p=2$) \cite{DeGennes:1993}. In contrast to these two cases, in which flow alignment occurs at arbitrarily small shear rates in the absence of confinement (and as long as $\lambda_{2}/\OPeq \ge 1$), for even $p>2$, flow alignment requires a high shear rate $\deps$ and is therefore expected to be non-universal. Specifically, at large shear rate, higher order powers of the strain rate can become comparable to those in Eq. (\ref{eq:hydrodynamics}c), thus affecting the magnitude of both $\deps_{\rm c}$ and $\theta_{p}$ in a system-dependent way. 

In conclusion, in this Letter we have revisited the hydrodynamic $p-$atic liquid crystals in two dimensions, with the goal of going beyond the classic theoretical picture based on continuous ${\rm O}(2)$ rotational symmetry. Our approach, build upon the $p-$atic tensor order parameter $\bm{Q}_{p}$, which directly embodies the discrete rotational symmetry of $p-$atic phases, allowed us to reveal novel couplings between $p-$atic order and flow, for which previous theories could not account. These couplings leave a distinct signature on the hydrodynamic of $p-$atics, such as the possibility of flow alignment at high shear rates, even for $p>2$. Furthermore, using fluctuating hydrodynamics, we have demonstrated that a shear flow of arbitrary finite shear rate has the remarkable effect of turning quasi-long-ranged orientational order, i.e. the hallmark of two-dimensional liquid crystals at equilibrium, into long-ranged order. Our theory could be experimentally tested on, e.g., free-standing liquid crystal films ~\cite{DeOliveira:2020,Jin:1996,Chou:1997,Chou:1998,Pindak:1998,Dierker:1986}. It could also serve as a starting point for the development of a hydrodynamic description of epithelial tissues, in light of the remarkable link between epithelia and hexatic liquid crystals established in Ref. \cite{Li:2018}. 

\acknowledgements 

We are indebted with Massimo Pica Ciamarra for insightful discussions. This work is partially supported by the ERC-CoG grant HexaTissue (L.G.) and by Netherlands Organization for Scientific Research (NWO/OCW), as part of the Vidi scheme (N.S. and L.G.) and the Frontiers of Nanoscience program (L.G.). JT thanks the Max-Planck Institut f\"ur Physik Komplexer Systeme,  Dresden, Germany for their hospitality, and their support through the Martin Gutzwiller Fellowship, and the Lorentz Center of the University of Leiden, Leiden, NL, for their support during a brief visit there,  while a portion of this work was underway.

\end{document}